\documentclass[12pt]{article}
\begin{document}

\title{Free Fermion Branches in some Quantum Spin  Models}
\author{F.~C.~Alcaraz$^{\rm a}$, Yu.~G.~Stroganov$^{\rm b,c}$\\
\small \it $^{\rm a}
$Universidade de S\~ao Paulo,
Instituto de F\'{\i}sica de S\~ao Carlos, \\
\small \it C.P. 369,13560-590, S\~ao Carlos, SP, Brazil \\
\small \it $^{\rm b}$Departamento de F\'{\i}sica, Universidade
Federal de S\~ao Carlos, 13565-905, \\
\small \it S\~ao Carlos, SP, Brazil \\
\small \it $^{\rm c}$Institute for High Energy Physics\\[-.5em]
\small \it 142284 Protvino, Moscow region, Russia}
\date{}

\maketitle

\begin{abstract}
Extensive numerical analysis of the eigenspectra of the $SU_q(N)$ invariant
Perk-Schultz Hamiltonian shows some simple regularities for a significant 
part of the eigenspectrum. Inspired by those results we have found two set of
solutions of the associated nested Bethe-ansatz equations. The first set 
is obtained at a special value of the anisotropy ($q = \exp(i2\pi (N-1)/N)$)
and describes in particular the ground state and nearby excitations 
as a sum of free-fermion quasienergies. The  second set of solutions provides 
the energies in the sectors whose number $n_i$ of particles of distinct 
species  ($i =0, \ldots, N-1$) are less or equal to the unity except for one 
of the species. For this last set we obtain the eigenspectra of a free fermion 
model for arbitrary values of the anisotropy. 
\end{abstract}

\vskip 1em

\begin{center}{\bf1. Introduction}
\end{center}

The nested Bethe ansatz is probably the most sophisticated algebraic construction 
of eigenvectors for integrable lattice models, where the underlying quantum group
 is of rank larger than one.
In the same way as  the usual Bethe ansatz, which was widely used during more 
then 60 years (see, e. g., \cite{rev} for reviews) 
 the amplitudes of the wave 
functions are expressed in terms of a sum of plane waves 
whose wave numbers are given in terms of non-linear and highly non 
trivial coupled equations known as the nested Bethe-ansatz equations (NBAE). 

Except in the thermodynamic limit, in very few cases the associated NBAE 
can be solved analytically. The exceptions happen for a small number of 
"spin waves" or for very small lattices. In addition to these, in the case 
where the equations are of non-nested type, as in the XXZ chain, analytical 
solutions for chains of arbitrary length  can also be derived for the 
free fermion point (anisotropy $\Delta = 0$) and for the special anisotropy 
$\Delta=-\frac{1}{2}$~\cite{alc1,YGS,ALL}.

In a recent paper~\cite{AS1} we presented a new set of analytical solutions 
of the NBAE for finite quantum chains.
These announced solutions correspond to the 
 $SU_q(3)$ Perk-Schultz model~\cite{PS}, 
or the  anisotropic $SU(3)$ Sutherland  model~\cite{Sutherland}, with 
periodic or open boundary conditions, at a special value of the anisotropy. 
These  solutions are given in terms of two types of wave numbers, however 
their number ($p_0,p_1$) do not satisfy the usual bounds ($p_0 <p_1<L$), and
 it is not clear if although being solutions of the NBAE  they 
do not 
correspond to zero-norm states. 
Moreover, even in the case where they correspond to physical eigenstates 
of the quantum Hamiltonian it is not clear, in general, 
 in what eigensector 
should we expect these solutions. In the case of periodic boundary 
conditions, by using the functional equations of the model we were able 
to answer partially the above points in \cite{AS1}.

In the present paper we fill the above deficiency by presenting a set of 
new analytical solutions of the NBAE having the  
number of wave numbers ($p_0,p_1$) satisfying 
the usual bounds $p_0<p_1<L$, and consequently explaining the free-fermion part of 
the eigenspectrum of the $SU(3)$ Perk-Schultz model conjectured from 
extensive numerical calculations in \cite{AS1}. Moreover our further "experimental" (i. e. 
numerical) investigations lead to new observations concerning the 
anisotropic $SU(N)$ Perk-Schultz quantum chains. 
Based on our numerical evidences we  again have formulated 
several conjectures. Surprisingly, the inspirations coming from these 
numerical observations enable us to find analytically a lot of solutions 
of the related NBAE, that explains part of these new conjectures. The list of
analytical solutions we described in this paper are:

\begin{itemize}
\item
the sectors $(n_0,n_1,n_2)=(k,k+1,L-2k-1)$ of the $SU(3)$ model with 
$q = \exp(2i\pi/3)$ with periodic boundary conditions.
\item
the sectors ($k,k,L-2k$) and ($k,k+1,L-2k-1$) of the $SU(3)$ model 
with $q = \exp(2i\pi/3)$ with free boundary conditions.
\item
the sector $(n_0,n_1,\ldots,n_{N-1}) = (1,1,\ldots,1,L-N+1)$ of the 
$SU(N)$ model with arbitrary values of $q$ and free boundary 
conditions.
\item
the sector $(n_0,n_1,\ldots,n_{N-1}) = (1,1,\ldots,1,2,L-N)$ of the 
$SU(N)$ model with $q=\exp((N-1)i\pi/N)$ and free boundary 
conditions. 
\item
the sector $(n_0,n_1,n_2,n_3) = (1,2,2,L-5)$ of the $SU(4)$ model 
with $q = \exp(i3\pi/4)$ and free boundary conditions
\footnote {We suspect that this class of solutions can be greatly generalized 
for the $SU(N)$ model with $q=\exp((N-1)i\pi/N)$  (see
discussion in the last section).}
\end{itemize}

%\end{itemize}

The paper is organized as follows.
In \S 2 we  introduce the model, and give the NBAE for periodic and free boundary conditions.
In \S 3 we consider   the set of analytical solutions related to 
the $SU(3)$  model with $q=\exp (2i\pi /3)$, with periodic and free boundary
conditions.
In \S 4  we consider the NBAE for the sector 
$(1,\ldots,1,L-N+1)$, with arbitrary values of the anisotropy.
We consider in details the $SU(4)$ case and formulate a mathematical 
statement that allows to  generalize our consideration to the case 
of arbitrary rank. Using a special case of this statement 
in \S 5  we find 
the NBAE free-fermion solutions  for the $SU(N)$ model with 
anisotropy  $q=\exp ((N-1)i\pi /N)$ in the eigensector 
($1,\ldots,1,2,L-N$).
\S 6 contains a set of new "experimental" conjectures that merged from
extensive numerical diagonalizations of the $SU(N)$ quantum chains for
arbitrary values of $N$. 
Finally  in \S 7 we present our conclusions and a summary of our results.
We also present in  
 appendix A and B the solutions   in the sector ($1,2,2,L-5$) of the $SU(4)$ 
model with anisotropy $q = \exp(i3\pi/4)$ and the wave functions related with  
the solutions considered in \S 4, respectively. 

\vskip 1em

\begin{center}{\bf2. The $SU(N)$ Perk-Schultz model}
\end{center}

The $SU(N)$ Perk-Schultz model~\cite{PS} is the anisotropic 
version of the 
$SU(N)$ Sutherland model~\cite{Sutherland} with the Hamiltonian, 
in a L-site chain, given by 

\begin{eqnarray}
\label{H}
&&H_{q}^{p} =\sum_{j=1}^{L-1} H_{j,j+1} + p H_{L,1}\qquad (p=0,1), 
\qquad \mbox{where}  \\ 
&&H_{i,j} =-\sum_{a=0}^{N-1} \sum_{b=a+1}^{N-1} 
( E_i^{ab} E_{j}^{ba} + E_i^{ba} E_{j}^{ab}- q E_i^{aa} E_{j}^{bb} 
- 1/q  E_i^{bb} E_j^{aa} ). \nonumber
\end{eqnarray}
The $N \times N$ matrices $E^{ab}$ have elements $(E^{ab})_{cd}=
\delta^a_c \delta^b_d$ and $q=\exp (i \eta)$ plays the role of  
the anisotropy of 
the model.
The cases of free and periodic boundary conditions are obtained 
by setting $p=0$ and $p=1$ in (\ref{H}), respectively. 
This Hamiltonian describe the dynamics of a system containing 
$N$ classes of particles ($0,1,...,N-1$) with on-site hard-core exclusion.
  The number of particles belonging to each specie is conserved separately.
 Consequently the 
 Hilbert space can be splitted into block disjoint sectors 
labeled by $(n_0,n_1,...,n_{N-1})$, where $n_i=0,1,...,L$ is the number 
of particles of specie i (i=0,1,...,N-1). The Hamiltonian (\ref{H}) has a $S_N$ 
symmetry due to its invariance under the permutation of distinct 
particles 
species, that implies that all the energies can be obtained from the 
sectors ($n_0,n_1,...,n_{N-1}$), where $n_0 \le n_1 \le ... \le n_{N-1}$ and  
$n_0+n_1+...+n_{N-1}=L$. 

At $q=1$ the model is $SU(N)$ invariant and for $q\neq 1$ the model has a
$U(1)^{\otimes {N-1}}$ symmetry as a consequence of the above mentioned 
conservation. Moreover in the special case of free boundaries ($p=0$), the
quantum chain (\ref{H}) has a larger symmetry, being invariant under the 
additional quantum $SU(N)_q$ symmetry. This last invariance implies 
that all the eigenenergies belonging to the sector 
($n_0^{\prime},n_1^{\prime},...,n_{N-1}^{\prime}$) with 
$n_0^{\prime} \le n_1^{\prime} \le ... \le n_{N-1}^{\prime}$  are degenerated 
with the energies belonging to the sectors 
$(n_0,n_1,...,n_{N-1})$ with $n_0 \le n_1 \le ... \le n_{N-1}$, 
if $n_0^{\prime} \le n_0$ and
$n_0^{\prime} + n_1^{\prime} \le n_0 +n_1$ and so on up to $n_0^{\prime} + 
n_1^{\prime}+\ldots +n_{N-2}^{\prime} \le n_0 +n_1+...n_{N-2}$.

 The NBAE that give  the eigenenergies of the  $SU_q(N)$ Perk-Schultz model 
in the sector whose number of particles is ($n_i, i =0,\ldots,N-1$) 
are given by (see e. g. ~\cite{resh},\cite{freeBAE})
\begin{eqnarray}
\label{ba3}
&&\prod_{j=1,j \ne i}^{p_k} F(u_i^{(k)},u_j^{(k)})= \prod_{j=1}^{p_{k-1}} 
f(u_i^{(k)},u_j^{(k-1)})  \prod_{j=1}^{p_{k+1}} f(u_i^{(k)},u_j^{(k+1)}),  \end{eqnarray}
where $k=0,1,...,N-2$ and $i=1,2...,p_k$.
The integer parameters $p_k$ are given by
\begin{eqnarray}
\label{pk}
&&p_k=\sum_{i=0}^{k} n_i, \quad k=0,1,...,N-2; \quad p_{-1}=0, \quad 
p_{N-1} = L.
\end{eqnarray}
The functions $F(x,y)$ and  $f(x,y)$ are defined by:
\begin{eqnarray}
\label{fper}
&&F(x,y)= \frac{\sin(x-y-\eta)}{\sin(x-y+\eta)},\quad   
f(x,y)= \frac{\sin(x-y-\eta/2)}{\sin(x-y+\eta/2)},
 \end{eqnarray}
for periodic boundary conditions,  and by 
\begin{equation}
\label{ff}
F(x,y)=\frac{\cos(2y)-\cos(2x-2\eta)}{\cos(2y)-\cos(2x+2\eta)},
 \quad   f(x,y)= \frac{\cos(2y)-\cos(2x-\eta)}{\cos(2y)-\cos(2x+\eta)},
 \end{equation}
for the free boundary case.
In using the  NBAE  (\ref{ba3}) we deal with variables of different order
$\{u_j^{\{k)}\}$ ($k=0,\ldots,N-2$).
The number of variables $u_{i}^{(k)}$ of order $k$ is equal to $p_k$.
The whole set of  NBAE consists of subsets of order $k$ which contain 
precisely $p_k$ equations ($k=0,1,...,N-2$).
 
The eigenenergies of the Hamiltonian (\ref{H}) in the sector ($n_0,n_1,...,n_{N-1}$) 
are given in terms of the roots $\{u_j^{(N-2)}\}$:
\begin{equation}
\label{ba1}
E=-\sum_{j=1}^{p_{N-2}} \biggl(-q-\frac{1}{q}+\frac{\sin(u_j-\eta /2)}
{\sin(u_j+\eta /2)} +\frac{\sin(u_j+\eta /2)}
{\sin(u_j-\eta /2)} \biggr),
\end{equation}
where to simplify the notation we write $u_j \equiv u_j^{(N-2)}$, 
($j=1,\ldots,p_{N-2}$).

All the solutions of the  NBAE (\ref{ba3}) which are going to be 
described in this paper satisfy the additional
"free-fermion" conditions (FFC):
\begin{eqnarray}
\label{ffc0}
&&f^L (u_i,0)=1 ,\quad i=1,...,p_{N-2}.
\end{eqnarray}
In this case, from (\ref{ff}) and (\ref{ba1}) the corresponding 
eigenenergies  of the Hamiltonian (\ref{H}) for the case of free 
boundaries are given by 
\begin{equation}
\label{ba1f}
E=-2\sum_{j=1}^{p_{N-2}} \biggl(-\cos \eta + \cos \frac{\pi k_j}{L}\biggr),\>\>1\le k_j\le L-1.
\end{equation}
On the other hand, for periodic boundaries we 
 have found solutions for the $SU(3)$ model with $\eta=2\pi/3$, and   for this case 
relations (\ref{fper}), (\ref{ba1}) and (\ref{ffc0}) give us 
 \begin{equation}
\label{ba1p}
E=-\sum_{j=1}^{p_{1}} \biggl(1 + 2 \cos \frac{2 \pi k_j}{L}\biggr),\>\>1\le k_j\le L.
\end{equation}

 With the additional set of equations merged from the FFC (\ref{ffc0}), the number 
of equations in (\ref{ba3}) and (\ref{ffc0}) exceeds the number of variables 
by $p_{N-2}$. 
 At the first glance we would expect  no chance to satisfy  the whole system 
given by (\ref{ba3}) and (\ref{ffc0}). 
But, surprisingly, the NBAE possess some hidden symmetry. 
In the next section, for the sake of simplicity,  we consider the simplest case of
 the $SU(3)$ model with $q=\exp (2i\pi /3)$. 

\vskip 1em

\begin{center}{\bf3. Free fermion spectrum for the  $SU(3)$ model with $q=\exp (2i\pi /3)$:    NBAE solutions.}
\end{center}

In \cite{AS1}, based on extensive numerical calculations we conjectured the
existence of free fermion solutions given by (\ref{ba1f}) and (\ref{ba1p}) for 
periodic and free boundary conditions, respectively. In the case 
of periodic boundaries these solutions happens in the sectors 
($n_0,n_1,n_2$) $=$ ($k,k+1,L-2k-1$) ($0\leq k\leq (L-1)/2$) while in the open
boundary case they belong to the sectors ($k,k,L-2k$) or ($k,k+1,L-2k-1$). 
In \cite{AS1} exploring the functional equations related to the NBAE we were 
able to explain partially these conjectures in the periodic case. 
In the free boundary case we derived directly the NBAE. These 
solutions however do 
 not belong to the numerically predicted 
sectors and do not satisfy the usual bounds $p_0 =n_0 < p_1 = n_0 + n_1 <L$. 
Moreover it is not clear if they correspond to non-zero norm eigenfunctions. 
In this section we present a direct explanation of these conjectures without 
the use of functional equations. 

The NBAEs  for the anisotropic $SU(3)$ Perk-Schultz model with anisotropy 
$q= \exp(2i\pi/3)$, with roots satisfying the FFC, can be written as 
follows: 
\begin{eqnarray}
\label{ba34}
&&\prod_{j=1,j \ne i}^{p_0} f(v_i,v_j) 
\prod_{j=1}^{p_1} f(v_i,u_j)=1, \quad i=1,2,\ldots,p_0,\\ 
\label{ba35}
&&\prod_{j=1,j \ne i}^{p_1} f(u_i,u_j) 
\prod_{j=1}^{p_0} f(u_i,v_j)=1,\quad i=1,2,\ldots,p_1, \\
\label{ffc1}
&&f^L (u_i,0)=1 ,\quad i=1,2,...,p_1,
\end{eqnarray}
where $p_0=n_0$,  $p_1=n_0+n_1$ and we have used the relation 
 \begin{eqnarray}
\label{uniq}
&& F(x,y)=1/f(x,y),
\end{eqnarray}
which is valid for $q=\exp(\frac{2i\pi}{3})$ or  $\eta=\frac{2\pi}{3}$.

\bf{3a. Periodic boundary condition}\rm
 
Before considering  the general case let us restrict ourselves initially   
to the 
particular eigensector $(1,2,L-3)$.  We have in this case $p_0=n_0=1$ and $p_1=n_0+n_1=3$.
For this simple case the first subsystem (\ref{ba34}) consists of a single  equation:
 \begin{eqnarray}
\label{spec12}
&& f(v,u_1) f(v,u_2) f(v,u_3)=1.
\end{eqnarray}
From the  definition of the function $f$ given in (\ref{fper}) 
we can show that the last equation (\ref{spec12}) is equivalent to 
  \begin{eqnarray}
\label{spec12fin}
&& \cos (v+u_1-u_2-u_3) + \cos (v-u_1+u_2+u_3) + \nonumber \\
&& + \cos (v-u_1-u_2+u_3) = 0.
\end{eqnarray}
It is clear that this relation has the $S_3$ symmetry under the permutation of the variables $u_1, u_2$ and $u_3$.
Surprisingly it has also the  $S_4$ symmetry under the permutation of the variables $u_1, u_2, u_3$ and $v$!

The three equations of the second subsystem (\ref{ba35})
can be obtained from (\ref{ba34}) by the permutations $v \leftrightarrow u_i$, 
so they also can be reduced to (\ref{spec12fin}).
Fixing the variables $u_i, \>\>i=1,2,3$,  satisfying the FFC (\ref{ffc1}) 
and finding $v$ from equation (\ref{spec12fin}) we  obtain the solution 
for the whole system  (\ref{ba34}-\ref{ffc1})
in the eigensector (1,2,L-3).

We show now that this  method can be generalized to any sector
($k,k+1,L-2k-1$),  supporting the conjecture 1 of our previous paper
\cite{AS1}. 
Let us consider the set $\{x_i, i =1,\ldots, p_0 + p_1\}$ of variables 
formed by the union of the  two systems of variables:  
\begin{eqnarray}
\label{xuv}
&&x_i=v_i\quad (i=1,2,\ldots ,p_0), \nonumber  \\
&&x_{i+p_0}=u_i\quad (i=1,2,\ldots ,p_1).
\end{eqnarray}
In terms of these variables the system of equations (\ref{ba34}-\ref{ffc1})
 becomes
\begin{eqnarray}
\label{ba36}
&&r(x_i)=f(x_i,x_i), \quad i=1,2,\ldots ,p_0+p_1,\\ 
\label{ffc2}
&& f^L (x_i,0)=1,\quad i=p_0+1,\ldots ,p_0+p_1,
\end{eqnarray}
where $r(x)=\prod_{j=1}^{p_0+p_1} f(x,x_j)$.
The first of these subsystems (\ref{ba36}) possess $S_{p_0+p_1}$ permutation symmetry.
We intend to show now that many of the equations in this subsystem 
are dependent, 
 and for $p_0=k$ and $p_1=2k+1$ we can also satisfy 
independently the   FFC (\ref{ffc2}).
From (\ref{fper}) we may write 
\begin{eqnarray}
\label{xb1}
&&r(x)=\prod_{j=1}^{p_0+p_1}\frac{\sin(x-x_j-\pi/3)}{\sin(x-x_j+\pi/3)}=
\prod_{j=1}^{p_0+p_1}\frac{b-q b_j}{q b-b_j},
\end{eqnarray}
where  for convenience we introduced the new variables 
\begin{equation}
\label{xb2}
b=\exp (2ix) ,\nonumber \quad
b_j=\exp (2ix_j), \quad
q=\exp (2i\pi /3). 
\end{equation}
The first subsystem  (\ref{ba36}) becomes now a system of algebraic equations:
\begin{eqnarray}
\label{ba38}
&&\prod_{j=1}^{p_0+p_1} (b_i-q\> b_j) + 
\prod_{j=1}^{p_0+p_1} (q\> b_i-b_j) = 0\quad i=1,2,\ldots ,p_0+p_1. 
\end{eqnarray}
Using  the standard symmetric functions:
\begin{eqnarray}
\label{symf}
&&S_0=1, \nonumber \\
&&S_1=b_1+b_2+\ldots +b_{p_0+p_1}, \nonumber \\
&&S_2=b_1\>b_2+b_1\>b_3+\ldots b_{p_0+p_1-1}\>b_{p_0+p_1}, \nonumber \\
&& \ldots  \\
&&S_m=\sum_{1\le i_1 < i_2 < \ldots i_m \le p_0+p_1} b_{i_1}\>b_{i_2}\>\ldots b_{i_m},  \nonumber \\
&& ... \nonumber \\
&&S_{p_0+p_1}=b_1\>b_2\>...b_{p_0+p_1}, \nonumber 
\end{eqnarray}
we can rewrite  (\ref{ba38}) as 
\begin{eqnarray}
\label{ba39}
&&\sum_{m=0}^{p_0+p_1} (-1)^m S_m b_i^{p_0+p_1-m} (q^m+q^{p_0+p_1-m})=0. 
\end{eqnarray}
Adding this last equation to the identity 
\begin{eqnarray}
\label{ide1}
&&q^{-p_0-p_1} \prod_{j=1}^{p_0+p_1} (b_i- b_j) = \sum_{m=0}^{p_0+p_1} 
(-1)^m S_m b_i^{p_0+p_1-m} q^{-p_0-p_1}=0, 
\end{eqnarray}
 we obtain 
\begin{eqnarray}
\label{ba40}
&&\sum_{m=0}^{p_0+p_1} (-1)^m S_m b_i^{p_0+p_1-m} (q^m+q^{p_0+p_1-m}+q^{-p_0-p_1})=0. 
\end{eqnarray}
For $q=\exp(2i \pi /3)$ we have the following possibilities 
\begin{eqnarray}
\label{qqq}
&&q^m+q^{p_0+p_1-m}+q^{-p_0-p_1}= 3q^{-p_0-p_1} \>\> \mbox{for} \>\> p_0+p_1+m=3n, \nonumber \\
 &&q^m+q^{p_0+p_1-m}+q^{-p_0-p_1}= 0\>\>  \mbox{for}\>\> p_0+p_1+m \ne 3n. 
  \end{eqnarray}
Let $p_0+p_1=3k+r$, where $k$ is an integer and $r \in \{-1,0,1\}$.
Inserting (\ref{qqq}) into (\ref{ba40}) we obtain
\begin{eqnarray}
\label{ba41}
&&\sum_{\mu=0}^{k} (-1)^{\mu} S_{3\mu} b_i^{-3\mu}=0  \>\> \mbox{for}\>\>r=0, \nonumber \\ 
&&\sum_{\mu=0}^{k-1} (-1)^{\mu} S_{3\mu+2} b_i^{-3\mu}=0 \>\> \mbox{for}\>\>r=1, \\ 
&&\sum_{\mu=0}^{k-1} (-1)^{\mu} S_{3\mu+1} b_i^{-3\mu}=0 \>\> \mbox{for}\>\>r=-1, \nonumber
\end{eqnarray}
where $i=1,2,...,p_0+p_1$. 
We then see that the  subsystem (\ref{ba36}) can be satisfied if we impose 
\begin{equation}
\label{s0}
S_{3\nu+\rho}=0,\>\>\nu=0,1,...,k-1,
\end{equation}
where $\rho=0$ for $r=0$, $\rho=2$ for $r=1$ and $\rho=1$ for $r=-1$.
 Since $S_0=1$ it is not possible to obtain solutions of this type 
for $L=3k$ ($\rho=0$) and we have to 
 limit ourselves to the cases where  $p_0+p_1 \ne 3k$.

Let us fix now $p_1$ variables $u_i$, satisfying the   FFC (\ref{ffc2}). 
Variables $v_i,\>i=1,2,...,p_0$ can, in principle, be found from the 
system (\ref{s0}).
In order to do that we use the 
decomposition of the symmetric functions depending on 
  two set of variables, namely  
\begin{eqnarray}
\label{dec}
&&S_0=1, \nonumber \\
&&S_1=s_1+\sigma _1, \nonumber \\
&&S_2=s_2+s_1 \sigma _1 + \sigma _2, \nonumber \\
&& ...  \\
&&S_m= \sum_{k=max\{0,m-p_0\}}^{min\{m,p_1\}} s_{m-k} \sigma _{k}, \nonumber \\
&& ... \nonumber \\
&&S_{p_0+p_1}= s_{p_0} \sigma _{p_1}, \nonumber 
\end{eqnarray}
where  $\sigma_i$ are the symmetric combination of the known variables $u_i$
and 
 $s_i$ are the symmetric combinations of the unknown variables
 $v_j$.
Consequently the 
 system (\ref{s0}) can be reduced to a linear system for symmetric functions $s_i,\>\>i=1,2,...,p_0$.
This system  can be solved 
if the number of variables $p_0$ is greater or equal to the number of equations $k$.   
Then  we have the system
\begin{eqnarray}
\label{lineq}
&&p_0=n_0, \quad p_1=n_0+n_1, \quad  n_0 \le n_1, \nonumber\\
&&p_0+p_1=3k \pm 1, \quad k \le p_0.
\end{eqnarray}
These relations give us the constraint $n_0=k$ and $n_1=k+1$, and consequently 
$p_0+p_1=2n_0+n_1 = 3k+1$, implying that  solutions of (\ref{s0}) exist  only 
for $\rho =2$. These  solutions happen in the sector ($k,k+1,L-2k-1$).
Consider for illustration the case $k=2$ ($p_0=2$ and $p_1=5$).
We have 5 variables $u_1,...,u_5$, which we fix with the FFC (\ref{ffc2}), 
and 2 unknown variables $v_1,v_2$.   
Using (\ref{dec}) we can write the  system (\ref{s0}) as follows:
 \begin{eqnarray}
\label{k=2}
&&S_2=s_2+s_1 \sigma _1 + \sigma _2 =0, \nonumber \\
&&S_5=s_5+s_4 \sigma _1 + s_3 \sigma _2 = 0 .
\end{eqnarray}
The functions $\sigma_1,\sigma_2,\sigma_3,\sigma_4$ and $\sigma_5$ we know, so we have 2 linear equations
for $s_1=v_1+v_2$ and $s_2=v_1 v_2$.

\bf{3a. Free boundary} condition \rm
 
The case of free boundary conditions is slightly more complicated.
Inserting  the definition (\ref{ff}) into the subystem (\ref{ba36})   we obtain  the system
\begin{eqnarray}
\label{baf1}
&&\sin(2x_i-2\pi/3) \prod_{j=1}^{p_0+p_1} (\cos(2x_j)-\cos(2x_i-2\pi/3))+\nonumber \\
&&+\sin(2x_i+2\pi/3) \prod_{j=1}^{p_0+p_1} (\cos(2x_j)-\cos(2x_i+2\pi/3))=0 \\
&& i=1,2,...p_0+p_1. \nonumber
\end{eqnarray}
As in the periodic case we may  use the symmetric functions:   
\begin{eqnarray}
\label{symff}
&&S_0=1, \nonumber \\
&&S_1=\cos 2x_1 + \cos 2x_2 + \cos 2x_{p_0+p_1}, \nonumber \\
&& ...  \\
&&S_{p_0+p_1}=\cos 2x_1\>\cos 2x_2\>...\cos 2x_{p_0+p_1}, \nonumber 
\end{eqnarray}
to  rewrite (\ref{baf1}) as 
\begin{eqnarray}
\label{baf2}
&&\sum_{m=0}^{p_0+p_1} S_{p_0+p_1-m} \biggl( \sin (2x_i-2\pi/3) \cos ^m (2x_i-2\pi/3))+\nonumber \\ 
&&+\sin (2x_i+2\pi/3) \cos ^m (2x_i+2\pi/3))\biggr) =0,  \\
&& i=1,2,...p_0+p_1. \nonumber
\end{eqnarray}
On  the other side we have the identities:
\begin{eqnarray}
\label{baf3}
&&\sin 2x_i\> \prod_{j=1}^{p_0+p_1} (\cos 2x_j - \cos 2x_i) = 0 
\nonumber \\ &&=\sum_{m=0}^{p_0+p_1} S_{p_0+p_1-m} \sin 2x_i \cos ^m 2x_i, \quad i=1,2,...p_0+p_1. 
\end{eqnarray}
Adding the equations  (\ref{baf3}) and  (\ref{baf4}) we obtain
\begin{eqnarray}
\label{baf4}
&&\sum_{m=0}^{p_0+p_1} S_{p_0+p_1-m} \phi _m(x_i)=0,\quad i=1,2,...p_0+p_1, 
\end{eqnarray}
where the functions $\phi _m (x) $ are given by 
\begin{eqnarray}
\label{baf5}
&&\phi _m (x)= \sum_{\rho =-1}^1 \sin (2x_i+2\pi \rho /3) \cos ^m (2x_i+2\pi \rho /3).  
\end{eqnarray}
It is convenient now to find a reccursion relation for these functions.
 In order to do that let 
 us consider a general sequence 
\begin{eqnarray}
\label{genseq}
&& y_m=a \alpha ^m + b \beta ^m +c \gamma ^m.
\end{eqnarray}
This sequence satisfy the recurrence relation  
\begin{eqnarray}
\label{recrel}
&& y_{m+1}+A y_m+B y_{m-1} + C y_{m-2} = 0,  
\end{eqnarray}
 if $A$, $B$ and $C$  are the coefficients of the third degree equation 
\begin{equation}
\label{3eq}
 t^3+A t^2+B t + C \equiv (t-\alpha) (t-\beta) (t-\gamma), 
\end{equation}
 with roots $\alpha$, $\beta$, and $\gamma$.
Consequently  our functions $\phi_m(x)$ satisfy the recurrence  relation:
\begin{equation}
\label{recphi}
 \phi_{m+1}(x)+A \phi_m(x)+B \phi_{m-1}(x) + C \phi_{m-2}(x) = 0, 
\end{equation}
where 
\begin{equation}
\label{3phieq}
 t^3+A t^2+B t + C \equiv \prod_{\rho=-1}^1 (t-\cos(2x+2\pi \rho/3))=
t^3-3/4\>t-1/4\cos 6x. 
\end{equation}
Due to  (\ref{recrel}) one then have  the recurrence relation  for the  
 functions $\phi_m$:
\begin{eqnarray}
\label{recphir}
&& \phi_{m+1}(x)= 3/4 \>\phi_{m-1}(x) + 1/4 \>\cos 6x \>\phi_{m-2}(x). 
\end{eqnarray}
In order to iterate (\ref{recphir}) we 
 need also $\phi_0$,$\phi_1$ and $\phi_2$. Using the definitions (\ref{baf5}) we find 
\begin{eqnarray}
\label{phi0}
&&\phi_0(x) =0, \quad \phi_1(x) =0, \quad \mbox{and} \quad \phi_2(x) =3/4\>\sin 6x.
 \end{eqnarray}
Let us give a list of several $\phi_i(x)$, which we obtain by using the initial 
functions (\ref{phi0}) and the recurrence relations  (\ref{recphir}): 
\begin{eqnarray}
\label{philist}
&&\phi_3(x)=0, \quad \phi_4(x) = 9/16\>\sin 6x , \quad \phi_5(x)=3/16\>\sin 6x \> \cos 6x,\nonumber \\
&&\phi_6(x)= 27/64\> \sin 6x,\quad
\phi_7(x)=9/32\>\sin 6x \> \cos 6x, \\
&&\phi_8(x) = \sin 6x \>(81/256 + 3/64\>\cos^2 6x).\nonumber
 \end{eqnarray}
It is clear from (\ref{phi0}) and (\ref{recphir}) that  $\phi_m(x)=\sin 6x \> 
\Phi_m(\cos 6x)$, where $\Phi_m(t)$ is a polynomial. The degree $\mu$ of this polynomial 
depend upon the number $m$. One can prove by finite induction that for  $m=3n+r,\>\>r=1,2,3$,  
$\phi_m =0$ if $n-2+r<0$ and otherwise it is a polynomial of degree
$\mu=n-2+r$.

Returning  to our main equation (\ref{baf4})
and using the value we just obtained for $\mu$ we easily verify  that the left side 
of this equation is (up to common factor $\sin 6x$) a polynomial of $\cos 6x$ with  degree
$\nu=[(p_0+p_1-2)/3]$,
where $[x]$ means the integer part of $x$.
We can try to  convert the  equation (\ref{baf4}) into an identity on the  variable $x$ by 
equating to 0 all coeffitients of the polynomial, i. e.: 
\begin{eqnarray}
\label{baf44}
&&\sum_{m=0}^{p_0+p_1} S_{p_0+p_1-m} \> \Phi _m(\cos 6x)=0.
\end{eqnarray}
We obtain $k=\nu+1=[(p_0+p_1+1)/3]$ linear equations on the symmetric combinations $S_m$. 
Before considering the general case let 
 us consider,  for the beginning, the   cases where $p_0+p_1=6,7$, and $8$.
Using  (\ref{philist}) the  relation among the  functions $\phi$ and $\Phi$ and 
the equation (\ref{baf44}), we obtain:
\begin{eqnarray}
\label{sn6}
&&S_4+3/4\>S_2+1/4\>S_1\>\cos 6x +9/16 = 0 \quad (p_0+p_1=6), \\
\label{sn7}
&& S_5+3/4\>S_3+1/4\>S_2\>\cos 6x +9/16\>S_1+ \nonumber \\
&&+3/8\>\cos 6x =0 \quad (p_0+p_1=7), \\
&& \nonumber \\
\label{sn8}
&& S_6+3/4\>S_4+1/4\>S_3\>\cos 6x +9/16\>S_2+\nonumber \\
&&+3/8\>S_1\>\cos 6x+27/64+1/16\>\cos^2 6x  =0\quad (p_0+p_1=8).
 \end{eqnarray}
We see that for $p_0+p_1=6$ and for  $p_0+p_1=7$ one has two equations:
\begin{eqnarray}
\label{6}
&&S_1=0, \nonumber \\
&&S_4+3/4\>S_2 +9/16 = 0,
\end{eqnarray}
and
\begin{eqnarray}
\label{7}
&&S_2+3/2=0, \nonumber \\
&&S_5+3/4\>S_3 +9/16\>S_1 = 0,
\end{eqnarray}
respectively.
As far as $p_0+p_1=8$ is concerned, the last term 
$1/16\>\cos^2 6x$ makes impossible to convert (\ref{baf44}) into an identity.

Let us consider  the general case. In the same way as in  the periodic 
boundary case we fix the variables 
$u_i$ ($i=1,\ldots,p_1$) satisfying the FFC
(\ref{ffc2}) and try to find the variables $v_i$ ($i=1,\ldots,p_0$) using 
$k=[(p_0+p_1+1)/3]$ equations, which we obtain by imposing that   all 
the coefficients of the polynomial (\ref{baf44}) are zero.
In order to find a solution  the number of equations $k$ can not 
 exceed the number of variables $p_0$.
 The whole restrictions 
\begin{eqnarray}
\label{lineqfree}
&&p_0=n_0, \quad p_1=n_0+n_1, \quad n_0 \le n_1,  \nonumber \\
&&p_0 \ge [(p_0+p_1+1)/3],
\end{eqnarray}
reduced to two variants:
$n_0=n_1=k,p_0+p_1=3k$ and $n_0=k,\>n_1=k+1,p_0+p_1=3k+1$, 
which correspond to the sectors $(k,k,L-2k)$ and
$(k,k+1,L-2k-1)$ respectively. 
One can show using the recurrence  relation (\ref{recphir}) 
and the initial functions (\ref{phi0}) that as long  $p_0+p_1\ne 3k+2$ 
we can obtain a consistent  linear system for the symmetric functions $S_i$.
 We use these linear equations  in the same way
as we have used  (\ref{s0}) in the periodic case and this explains 
the conjecture 3 of \cite{AS1}. 
For example,  the  systems 
(\ref{6}) and (\ref{7}) give us free-fermion 
solutions for the sectors $(2,2,L-4)$ and $(2,3,L-5)$.
 
\vskip 1em

 \begin{center}{\bf4. Free fermion solutions of the NBAE  for generic $q$}
\end{center}

In this section, differently from the results of last section, 
where the free fermion solutions were found for a specific value of $q$, 
 we are 
going to present free fermion solutions of the NBAE that are valid 
for arbitrary values of $q$. These solutions are valid 
only for free boundary conditions and will happen in the special 
sectors of the $SU(N)_q$ model where the number of particles of each 
distinct specie is at most one, except for one of the species, that can be
considered as the background (holes for example).
The $S(U)_q$ symmetry ensures that the general sector containing all these 
solutions is the special sector where ($n_0,\ldots,n_{N-1}$) $=$ 
($1,\ldots,1,L-N+1$), that gives from the definition (\ref{pk}) the 
values 
\begin{eqnarray}
\label{pk0}
&&p_0=1, p_1 = 2, ... , p_{N-2}=N-1, p_{N-1}=L.
\end{eqnarray}
For this case the NBAE (\ref{ba3}) consist of $N-1$ subsets ($k=0,1,\ldots,N-1$) 
of equations where the $k$th subset has precisely $k+1$ equations.   

To illustrate our procedure  let us consider for simplicity the $SU(4)$ model  
 in the sector ($1,1,1,L-3$). In this case we have three subsets ($k=0,1,2$) 
 and three groups of variables ($u^{(0)},u^{(1)},u^{(2)}$):
\begin{eqnarray}
\label{su41}
&& 1=  f(w,v_1) f(w,v_2), \\
&& \nonumber \\
\label{su42}
&&F(v_1,v_2)= f(v_1,w) \times f(v_1,u_1) f(v_1,u_2) f(v_1,u_3),\nonumber \\
&&F(v_2,v_1)= f(v_2,w)  \times  f(v_2,u_1) f(v_2,u_2) f(v_2,u_3),  \\
&& \nonumber  \\
\label{su43}
&&F(u_1,u_2) F(u_1,u_3)= f(u_1,v_1)\>f(u_1,v_2),   \nonumber \\
&&F(u_2,u_1) F(u_2,u_3)= f(u_2,v_1)\>f(u_2,v_2),    \\
&&F(u_3,u_1) F(u_3,u_2)= f(u_3,v_1)\>f(u_3,v_2),   \nonumber  
\end{eqnarray}
where we write $w$ instead of $u^{(0)}$, $v_i$ instead of $u_i^{(1)}$ 
($i=1,2$) and 
$u_i$ instead of $u_i^{(2)}$ ($i=1,2,3$).
Inserting (\ref{ff}) into the first equation (\ref{su41}) we obtain promptly
  two possibilities:
\begin{eqnarray}
\label{v=0}
&& \sin 2w =0,  \\
\label{main}
&&\cos 2 v_1 + \cos 2 v_2 - 2 \cos \eta \cos 2w =0.
\end{eqnarray}

We do not intend to consider for the moment the first possibility  (\ref{v=0}).
Imposing the condition (\ref{main}) we can check (eliminating, for example 
the variable $w$) that for ($v_1,v_2,w$) satisfying (\ref{main}), 
besides (\ref{su41}) we have also two additional equalities, namely 
\begin{eqnarray}
\label{su3res}
&&F(v_1,v_2)= f(v_1,w), \nonumber \\
&&F(v_2,v_1)= f(v_2,w).
\end{eqnarray}
Consequently  the second subsystem become:
\begin{eqnarray}
\label{su4sec}
&&1 = f(v_1,u_1)\>f(v_1,u_2)\>f(v_1,u_3), \nonumber \\
&&1 = f(v_2,u_1)\>f(v_2,u_2)\>f(v_2,u_3). 
\end{eqnarray}
We will show bellow that the set of  5 equations formed by  the second (\ref{su4sec}) and 
the third subsets (\ref{su43})  contain only two independent equations, and 
consequently we may fix the variables  $u_i (i=1,2,3)$ by imposing the 
FFC (\ref{ffc2} and find the two variables $v_1$ and $v_2$ from these equations. 
The remaining variable $w$ is then obtained from  (\ref{main}). 
Consequently we find the free-fermion eigenspectra (\ref{ba1f}) for 
arbitrary values of $q$ or $\eta$. 

The previous analysis can be extended for the sector 
($1,\ldots,1,L-N+1$) for general $SU(N)_q$ by exploring the 
general theorem:

{\it For any $k$ fixed numbers $u_1,u_2,... u_k$ one can find $k-1$ numbers 
$v_1,v_2,..., v_{k-1}$ satisfying the two set of equations} 
\begin{eqnarray}
\label{maint1}
&&  \prod_{j=1}^k f(v_i,u_j) =1 \quad (i=1,2,...,k-1),\\
\label{maint2}
&& \prod_{j=1,j \ne i}^k F(u_i,u_j) =  \prod_{j=1}^{k-1} f(u_i,v_j) \quad (i=1,2,...,k).
\end{eqnarray}

Leaving the proof of the above theorem for the moment, we can see that applying
this theorem to  ($k-1$) known numbers $v_1,\ldots,v_{k-1}$, one can find 
the numbers $w_1,\ldots,w_{k-2}$ satisfying the equations 
\begin{eqnarray}
\label{maint3}
&&  \prod_{j=1}^{k-1} f(w_i,v_j) =1 \quad (i=1,2,...,k-2), \\
\label{maint4}
&& \prod_{j=1,j \ne i}^{k-1} F(v_i,v_j) =  \prod_{j=1}^{k-2} f(v_i,w_j) \quad (i=1,2,...,k-1).
\end{eqnarray}
Multiplying the $i$th ($i=1,\ldots,k-1$) equation of the sets (\ref{maint1}) and 
(\ref{maint4}) we obtain  literally one of the subset of the NBAE:
\begin{eqnarray}
\label{maintr}
&& \prod_{j=1,j \ne i}^{k-1} F(v_i,v_j) =  \prod_{j=1}^{k-2} f(v_i,w_j)  \prod_{j=1}^k f(v_i,u_j) \quad (i=1,2,...,k).
\end{eqnarray}

Applying the above theorem $k-1$ times we obtain a tower  of numbers:
\begin{eqnarray}
\label{pyr}
&& u_1,u_2,u_3......u_{k-2},u_{k-1},u_k \nonumber \\
&& \>\>v_1,v_2.....v_{k-2},v_{k-1} \nonumber \\
&&\>\>\>\> w_1,...w_{k-2} \nonumber \\
&& \>\>\>\>\>\>............\nonumber \\
&& \>\>\>\>\>\>\>\>......... \nonumber \\
&& \>\>\>\>\>\>\>\>y_1,y_2 \nonumber  \\
&& \>\>\>\>\>\>\>\>\>z \nonumber 
\end{eqnarray}
This imply that if we begin by fixing the $k=N-1$ variables 
$u_i$ ($i=1,\ldots,k$), satisfying the FFC (\ref{ffc2})
we obtain by using recursively the construction (\ref{maintr}) the solution of 
the NBAE of the $SU(N)$ Perk-Schultz model, with free boundaries, in the 
sector 
 $(n_0,n_1,\ldots,n_{N-1}) = (1,1,\ldots,1,L-N+1)$. 
The free-fermion like energies are given by (\ref{ba1f}) for arbitrary values 
of $\eta$. The previous results 
(\ref{su43}) and (\ref{su4sec}) are just consequences of the 
particular case where $k=3$.

Let us now prove the announced theorem.
Let us fix  $\{u_j\}, j=1,2,...,k$.
The equation (\ref{maint1}) can then be written   as follows
\begin{equation}
\label{Pvi}
 P(v_i)=0,\quad (i=1,2,...,k)
\end{equation}
where
\begin{equation}
\label{Pv}
 P(v) \equiv \prod_{j=1}^{k} \biggl ( \cos 2u_j -\cos (2v-\eta)\biggr) - \prod_{j=1}^{k} \biggl( \cos 2u_j -\cos (2v+\eta)\biggr).
\end{equation}
The use of  the symmetric functions (\ref{symff})  ($u_j$ instead of $x_i$) 
allow us to write 
\begin{equation}
\label{Pv1}
 P(v) \equiv \sum_{m=1}^k (-1)^{m+1} S_{k-m} \biggl ( \cos^m (2v-\eta) -\cos^m (2v+\eta)\biggr).
\end{equation}
Since  $\cos (2v-\eta) -\cos (2v+\eta)= 2 
\sin \eta \sin 2v$ and $a^m-b^m = (a-b) (a^{m-1}+a^{m-2} b +...+b^{m-1})$  we have 
\begin{equation}
\label{Pv2}
 P(v) \equiv \sin 2v\>p(\cos 2v),
\end{equation}
where $p(t)$ is a polynomial of degree $k-1$.
We can factorize this polynomial, and apart form  a multiplicative 
constant ($\Omega$)we can write 
\begin{equation}
\label{Pv3}
P(v) = \Omega \sin 2v\>\prod_{i=1}^{k-1}(\cos 2v - \cos 2b_i).
\end{equation}
Now (\ref{Pvi}) is easily solved: $v_i=b_i,\>\>i=1,2,...,k-1$.

Consider now the right side of the second equation (\ref{maint2}). 
The relation (\ref{Pv3}) allow us to write 
\begin{eqnarray}
\label{Pv4}
&& \prod_{i=1}^{k-1} f(u_j,v_i)=\prod_{i=1}^{k-1} f(u_j,b_i)=
\prod_{i=1}^{k-1} \frac{\cos(2u_i-\eta)-\cos2 b_i}{\cos(2u_i+\eta)-\cos2 b_i}=\nonumber \\
&& = \frac{P(2u_j-\eta)}{\sin(2u_j-\eta)} \frac{\sin(2u_j+\eta)}{P(2u_j+\eta)}. \end{eqnarray}
Using the expression (\ref{Pv}) for $P(v)$ we obtain:
\begin{eqnarray}
\label{Pv5}
&& \prod_{i=1}^{k-1} f(u_j,v_i)=\frac{\sin(2u_j+\eta)}{\sin(2u_j-\eta)}  \nonumber \\
&&\times \frac{\prod_{i=1}^{k} ( \cos 2u_i -\cos (2u_j-2\eta)) - \prod_{i=1}^{k} ( \cos 2u_i -\cos 2u_j)}
{\prod_{i=1}^{k} ( \cos 2u_i -\cos 2u_j) - \prod_{i=1}^{k} ( \cos 2u_i -\cos (2u_j+2\eta))}.
\end{eqnarray}
Since  $\prod_{i=1}^{k} ( \cos 2u_i -\cos 2u_j) =0$,  we obtain
\begin{equation}
\label{Pv6}
 \prod_{i=1}^{k-1} f(u_j,v_i)=-\frac{\sin(2u_j+\eta)}{\sin(2u_j-\eta)} 
 \prod_{i=1}^{k} \frac{ \cos 2u_i -\cos (2u_j-2\eta)} 
{  \cos 2u_i -\cos (2u_j+2\eta)}.
\end{equation}
One can also check that 
\begin{eqnarray}
\label{Pv7}
\frac{\sin(2u_j+\eta)}{\sin(2u_j-\eta)}=-\frac{ \cos 2u_j -\cos (2u_j+2\eta)} 
{  \cos 2u_j -\cos (2u_j-2\eta)},
\end{eqnarray}
so that (\ref{Pv6}) can be written as
\begin{eqnarray}
\label{Pvfin}
&& \prod_{i=1}^{k-1} f(u_j,v_i)= \prod_{i=1,i \ne j}^{k} F(u_j,u_i),
\end{eqnarray}
concluding the proof of the theorem.

In Appendix B  we derived the wave functions corresponding to the free 
fermion solutions obtained in this section.

 \begin{center}{\bf5. Free fermion spectrum for  special 
values of $q$. 
NBAE solutions for the  sector ($1,1,\ldots,1,2,L-N$)}
\end{center}

For this sector the NBAE of the Hamiltonian (\ref{H}) with free 
boundaries  have about the same system of equations 
 as in the previous section but 
the number of variables $u_i$ is now  $k+1$ instead of $k$.   
We have in this case to use a  modified theorem:

{\it For any $k+1$ fixed numbers $u_1,u_2,... u_k,u_{k+1}$ one can find $k-1$ numbers 
$v_1,v_2,..., v_{k-1}$ so that the two systems of equations} 
\begin{eqnarray}
\label{maint1m}
&&  \prod_{j=1}^{k+1} f(v_i,u_j) =1 \quad (i=1,2,...,k-1),\\
\label{maint2m}
&& \prod_{j=1,j \ne i}^{k+1} F(u_i,u_j) =  \prod_{j=1}^{k-1} f(u_i,v_j) \quad (i=1,2,...,k),
\end{eqnarray}
{\it are satisfied for the special anisotropy $\eta=k \pi /(k+1)$}.

We begin the  proof as in the case of previous theorem 
and as in (\ref{Pv2}) the corresponding polynomial $P(v)$ 
we obtain would have a degree $k$, however for the special  anisotropy value 
$\eta = k\pi/(k+1)$ we can show that the 
degree of $P(v)$ decreases to $k-1$, and we have the same situation as in 
the previous section. Consequently the proof follows straightforwardly in 
the same way as before and we have for the sector ($1,\ldots,1,2,L-N$) the free 
fermion energies given by (\ref{ba1f}), with $p_{N-2} = N$ and 
$\eta = (N-1)\pi/N$.

\vskip 1em

\begin{center}{\bf6. Free fermion spectrum for $SU(N)$ model.
Conjectures merged from numerical studies}
\end{center}

In this section we report some conjectures 
 merged from extensive brute-force 
numerical diagonalizations of the quantum chain (\ref{H}) with 
free boundary condition The analytical results presented in previous sections 
explain part of these numerical observations. In order to announce these 
conjectures let us  
 define the special sectors
\begin{equation}
\label{sectors}
S_k=([\frac{k}{N-1}],[\frac{k+1}{N-1}],...,[\frac{k+N-2}{N-1}],L-k), \qquad k=0,1,...,L,
\end{equation}
where, as before,  $[x]$ means the integer part of $x$.
For example, for  $N=4$ and $L=7$ the sectors are
\begin{eqnarray}
\label{s7}
&&S_0=(0,0,0,7), \quad 
S_1=(0,0,1,6), \quad 
S_2=(0,1,1,5), \nonumber \\
&&S_3=(1,1,1,4), \quad
S_4=(1,1,2,3), \quad
S_5=(1,2,2,2), \nonumber \\
&&S_6=(2,2,2,1), \quad 
S_7=(2,2,3,0),  
\end{eqnarray}
and due to the quantum symmetry of the 
Hamiltonian we have the  special ordering:
\begin{equation}
\label{order}
S_0 \subset S_1 \subset S_2 \subset S_3 \subset S_4 \subset S_5 \equiv  S_6 \supset S_7.
\end{equation}
This means that, for example,  all eigenvalues found in the sector $S_3$ can also 
be found  in the sectors $S_4,S_5$ and $S_6$, and  all eigenvalues appearing 
in the  sector $S_7$ can also be found in  the sectors $S_6$ and $S_5$.
The sectors $S_5$ and $S_6$ are totally  equivalent.
In this example let  us call the sectors $S_0,S_1,S_2,S_3,S_4,S_5$ as the {\it left} sectors 
and $S_6,S_7$ as the {\it right} ones.

We can  generalize this definition to any $L= N n + r $, where
n and r are natural numbers  and $r<N$,
obtaining $L-n$ left sectors and $n+1$ right ones.
Now we can formulate 
 the conjectures merged form our 
 bruteforce diagonalizations.  

CONJECTURE 1. {\it For $L=N n+r$ ($r=1,2,\ldots,N-1$) the Hamiltonian (\ref{H}) with $p=0$ and $q=\exp(i \pi/N)$ 
has eigenvalues given  by }
\begin{equation}
\label{Efree}
E_{I} = -2 \sum_{j\in I} (\cos(\frac{\pi}{N}) + \cos(\frac{j \pi}{L})),
\end{equation}
{\it where $I$ is an arbitrary subset of the set $\{1,2,...L-1\}$.
If  $k$ is the number of elements of the subset $I$ and also  $S_k$ is a left sector 
 then we find the eigenvalues (\ref{Efree}) in the sectors $S_k,S_{k+1},...S_{L-n}$.
On the other hand if  $S_k$ is a right sector then we find the eigenvalues (\ref{Efree}) 
in the sectors $S_{L-n-1},S_{L-n},\ldots,S_{k+1}$.}

For $L=N n$ we have slightly more delicate picture.
In this case we have the left sectors, the right sectors and a central one
($n,n,n,n$).

CONJECTURE $1^{\prime}$. {\it For $L=N n$ we can use conjecture 1 considering  
the central sector as a left or a right one.} 
This  is possible due to the  coincidence, apart of degeneracies,  of the 
eigenenergies in   the eigensectors 
($n,n,n,\ldots,n$) and ($n-1,n,n,\ldots,n+1$).

Consider now the a special subsets $I=\{1,2,3,\ldots,k\}$,  
$k=0,1,\ldots,L-1$.
Due to the conjectures 1 and $1^{\prime}$ the Hamiltonian (\ref{H}) has 
the corresponding eigenvalues
\begin{equation}
\label{Ek}
E^{(k)} = -2\sum_{j=1}^{k} (\cos(\frac{\pi}{N})+\cos(\frac{j \pi}{L})) 
= 1-2 k \cos(\frac{\pi}{N})-\frac{\sin \pi (2k+1)/2L}{\sin \pi/2L}.
\end{equation}
We can  now  formulate the remarkable conjecture:

CONJECTURE 2. {\it The eigenvalues  (\ref{Ek}) are the lowest  
eigenenergies  in the special sectors (\ref{sectors}) of the  
Hamiltonian (\ref{H}) with anisotropy $q=\exp(i(N-1) \pi/N)$.
 Namely, for the  left sectors we have $E_{min}(S_k)=E^{(k)}$ 
while for the right sectors 
$E_{min}(S_k)=E^{(k-1)}$.}

\noindent  We can add here an additional conjecture: 

CONJECTURE $2^{\prime}$. {\it The eigenvalues  (\ref{Ek}) for $k=L-n-1$ 
gives the ground state energy  of the Hamiltonian: }
\begin{equation}
\label{Eminfree}
E_0 = 1 + 2 (1-L+n) \cos(\frac{\pi}{N}) -\frac{\sin \pi (2n+1)/2L}{\sin \pi/2L}.
\end{equation}
%Remind that $n=[L/N]$

The last of the special sectors is 
\begin{equation}
\label{lastsec}
S_L=([\frac{L}{N-1}],[\frac{L+1}{N-1}],...,[\frac{L+N-2}{N-1}],0), \qquad k=0,1,...,L.
\end{equation}
In this sector we have only $N-1$ classes of particles and the   Hamiltonian
(\ref{H}) is reduced effectively to a $SU_q(N-1)$-invariant quantum spin model.
The sector is of right type and due to conjecture 2 we can state the following 
conjecture: 

CONJECTURE 3.  {\it The eigenvalue  (\ref{Ek}) for $k=L-1$ gives the ground state energy of 
the $SU_q(N-1)$ Hamiltonian with anisotropy $q=\exp(i(N-1)\pi/N)$. 
This eigenvalue can be written as } 
\begin{equation}
\label{Eminpred}
E_0 = - 2 (L-1) \cos(\frac{\pi}{N}).
\end{equation}

For $N=3$, for example,  we get the XXZ model with anisotropy $\Delta=-1/2$ and $E_0=1-L$, 
a result that was first observed in \cite{alc1} and produced quite interesting 
consequences \cite{YGS,ALL}.

\begin{center}{\bf7. Summary and conclusions}
\end{center}

Most of the researches  related to exact integrable systems ends with the 
derivation of the Bethe-ansatz equations, whose solutions provide, in 
principle, the eigenvalues and eigenvectors of the Hamiltonian or 
transfer matrix of the associated model. These equations are highly non linear 
and in several cases thanks to some appropriate guessing on the topology of 
the roots in the complex plane, they are transformed in the bulk limit 
($L \rightarrow \infty$) into integral equations, that provide most of the 
thermodynamic physical quantities, whose calculations depend upon 
eigenvalues only. Even in this limit theses simplifications does not help for the 
calculation of quantities depending directly on the eigenvectors like form 
factors and correlation functions, since the amplitudes of the eigenvectors, 
obtained in the Bethe basis, are given as combinatorial sums of plane waves 
that are highly nontrivial to operate in the thermodynamic limit. 

Since the exact integrability is a lattice size independent 
property, 
the exact solution of the associated Bethe-ansatz equations is a  quite 
important step towards the complete mathematical and physical understanding 
of these models. Due to the complexity of these equations, the solutions 
are known analytically in very few particular cases, like the free 
fermion case and the XXZ chain with anisotropy $\Delta = -1/2$ \cite{alc1}.
It is interesting to mention that in this last case only some ratio 
among the amplitudes of the eigenvectors are known \cite{YGS} and 
these results provide a quite interesting connection of this problem 
with that of the sign-alternating matrices \cite{ALL}.  In a previous paper 
\cite{AS1} we presented some new solutions of the associated Bethe ansatz 
equations of the simplest generalization of the XXZ chain with two 
conserved global quantities, namely the $SU(3)_q$ Perk-Schultz 
model with periodic  and free boundary conditions. 
In this case the Bethe ansatz equations are of nested type (NBAE) and these 
solutions were obtained for the special anisotropy value $q = 
\exp(i2\pi/3)$, being the first example of analytical solutions of NBAE for 
finite lattices. The results in \cite{AS1} were derived by exploring the 
functional relations merged from the associated NBAE, and presented 
two difficulties: the solutions for the open lattice happen in sectors out of
the usual bounds ($n_0 <n_1<n_2$), being not clear if they do not correspond 
to zero-norm states, and the precise sector where the periodic solutions 
happen is not known. In \S 3 of the present paper, in a different way, by
adding to the set of NBAE the FFC (\ref{ffc1}) we were able to solve these
difficulties by showing the existence of free-fermion solutions in the sectors
satisfying the usual bounds ($n_0 <n_1<n_2$) for the periodic and open cases.
On the contrary to the functional method used in \cite{AS1}, the success of the
direct method presented in \S 3 for the $SU(3)_q$ case enable us to generalize
these solutions for some sectors of the Perk-Schultz $SU(N)_q$ models,
 revealing the existence of free-fermion
branches, at the special anisotropy $q = \exp(i2\pi(N-1)/N)$. It is important to
mention that our numerical calculations reveal other energies beyond those
predicted in the calculated free-fermion branches. Although we did not prove
analytically, an extensive numerical analysis shows that the low lying energies,
including the ground-state energy belong to these classes of free-fermion like
solutions we have found. In fact it was these numerical results that induce our
analytical calculations. The numerical analysis indicated a set of conjectures
presented in \S 6, part of them were proved in \S 3 - \S 5 and appendix A. 

After the success of finding free-fermion solutions at special values of
anisotropies of the $SU(N)_q$ model, we tried to find other solutions for 
other anisotropies, at least for some simple sectors. Surprisingly we were 
able to find free-fermion solutions for all values of the anisotropy in the
sector were we have at most one particle of each specie, except for one of the
species (the background one). In \S 4 we show the existence of these solutions
for the most general sector within  this class, i. e., 
($n_0,\ldots,n_{N-1}$) $=$ ($1,\ldots,1,L-N+1$). Beyond the free-fermion
energies degenerated with the lower sectors, appear in this sector the energies
given by the sum of $p_{N-2} =N-1$ free-fermion energies $\epsilon_j$, i. e., 
\begin{equation} \label{conc1}
E = \sum_{j=1}^{N-1} \epsilon_j = \sum_{j=1}^{N-1} (q +\frac{1}{q} -x_j 
-\frac{1}{x_j}),
\end{equation}
where
\begin{equation} \label{conc2}
x_j = \exp(i\frac{\pi k_j}{L}), \quad 1\leq k_j \leq L-1.
\end{equation}
These eigenvalues expressions, valid for arbitrary values of $q$, induce us to
expect that a direct analytical derivation of the related eigenfunctions,
without the use of the NBAE might  be possible.  In fact in the appendix B we
derive these eigenvectors, which are given by combinations of Slater
fermionic determinants of one particle wave functions $\Psi_j(m_l)$. 
Denoting the configurations where the $l$th particle is at position $m_l$ 
($l=1,\ldots,N-1; m_l=1,\ldots,L$) as $|m_1,\ldots,m_{N-1}>$ the
eigenfunctions are given by 
\begin{eqnarray} \label{conc3}
&&|\psi_{\{x_1,\ldots,x_{N-1}\}}> = \sum_{m_1,\ldots,m_{N-1}= 1}^L
q^{-f(m_1,\ldots,m_{N-1})} \nonumber \\
&& \times \mbox{det}
\left|\begin{array}{cccc} 
\Psi_1(m_1) & \Psi_1(m_2) & \cdots & \Psi_1(m_{N-1}) \\
\Psi_2(m_1) & \Psi_2(m_2) & \cdots & \Psi_2(m_{N-1}) \\
\cdots      & \cdots      & \cdots & \cdots          \\
\Psi_{N-1}(m_1) & \Psi_{N-1}(m_2) & \cdots & \Psi_{N-1}(m_{N-1}) 
\end{array} \right|
|m_1,\ldots,m_{N-1}>, \nonumber
\end{eqnarray}
where 
\begin{equation} \label{conc4}
\Psi_j(m) = \left( 1-\frac{q}{x_j}\right)x_j^m - \left(1-qx_j\right)/x_j^m.
\end{equation}
The factor $f(m_1,\ldots,m_{N-1})$ is an integer given by the minimum number 
of pair permutations of the set $(m_1,\ldots,m_{N-1}) \rightarrow 
(m_1',\ldots,m_{N-1}')$, such that 
$m_1' <m_2'< \cdots <m_{N-1}'$. It is interesting to observe that at the isotropic 
point the wave function amplitudes are just given by standard free fermion 
Slater determinants.

In \S 5,  continuing  our search for free-fermion like solutions we were able 
again to find solutions in the sector ($n_0,\ldots,n_{N-1}$) $=$ 
($1,\ldots,1,2,L-N$). In this case the free fermion solutions 
happens only at the special value of the anisotropy $\eta= (N-1)\pi/N$. 
 Although we believe that should be possible to obtain free-fermion like 
solutions for arbitrary eigensectors, this procedure is not straightforward, and each case 
deserving some ingeniousness. As another example we show in appendix A the existence 
of free fermion solutions in the sector 
($n_0,n_1,n_2,n_3$) $=$ ($1,2,2,L-5$) of the $SU(4)_q$ model at the 
anisotropy $q = \exp(i3\pi/4)$.

Although we have only derived wave functions for the sector ($n_0,\ldots,n_{N-1}$) 
$=$ ($1,\ldots,L-N+1$) we expect to obtain in the future the 
corresponding eigenfunctions for the other sectors. This program is very 
important, since for example, it may teach us to derive in a closed  form the 
wave function for the interesting problem of the $\Delta = -1/2$ 
XXZ chain, or the sector ($0,n,L-n$) ($n=0,\ldots,L$) of the $SU(3)_q$ 
model at $q=\exp(i2\pi/3)$. 

\vspace{1.5cm}

{\it Acknowledgments}  This work was supported 
in part by the brazilian agencies FAPESP and CNPQ (Brazil), and by the Grant 
 \# 01--01--00201 (Russia) and INTAS 00-00561.

\begin{center}{\bf Appendix A. SU(4) model with $q=\exp (3i\pi /4)$,\\ sector $(1,2,2,L-5)$.}
\end{center}

The nested Bethe ansatz for $SU(4)$ Perk-Schultz model with free boundary
can be written as follows (see (\ref{ba3}) and (\ref{ff})):

\begin{eqnarray}
\label{apbae1}
&&\prod_{j=1,j \ne i}^{p_0} F(w_i,w_j)= \prod_{j=1}^{p_1} 
f(w_i,v_j) ,\>\> i=1,2,...,p_0, \\ 
\label{apbae2}
&&\prod_{j=1,j \ne i}^{p_1} F(v_i,v_j)= \prod_{j=1}^{p_0} 
f(v_i,w_j)  \prod_{j=1}^{p_2} f(v_i,u_j), \>\>i=1,2,...,p_1, \\ 
\label{apbae3}
&&\prod_{j=1,j \ne i}^{p_2} F(u_i,u_j)= \prod_{j=1}^{p_1} 
f(u_i,v_j), \>\> i=1,2,...,p_2  \\
\label{apffc}
&&f^L(u_i,0)=1,\>\>  i=1,2,...,p_2, 
\end{eqnarray}
where we restricted ourselves to the  solutions 
satisfying the FFC (\ref{ffc0}). 

We have found free-fermion solutions for arbitrary values of $q$ in the sector 
($n_0,n_1,n_2,n_3$) $=$ ($1,1,1,L-3$) (see \S 4) and for 
$q=\exp(3i\pi/4)$ in the sector ($1,1,2,L-4$) (see \S 5). We 
intend now to obtain solutions in the next simple sector
($1,2,2,L-5$).
 For this case  $p_0=1$, $p_1=3$, $p_2=5$. Choosing  the five variables $u_i$ satisfying the FFC (\ref{apffc}),  we look for three variables
$v_i$ and one variable $w$, satisfying the  nine  equations given in 
(\ref{apbae1})-(\ref{apbae3}).    

Let us begin with the third group of equations (\ref{apbae3}).  
The left side of these equations  can be written as
 $L(u_i),\>\>i=1,\ldots,5$, where 
 \begin{eqnarray}
\label{left1}
&&L(u)=\frac{1}{F(u,u)} \prod_{j=1}^{5} F(u,u_j) 
\end{eqnarray}
Using the definition (\ref{ff}) and fixing the anisotropy  
parameter $\eta=3\pi/4$
we obtain after tedious but straightforward calculations:
\begin{eqnarray}
\label{left2}
&&\frac{L(u)-1}{L(u)+1}=s \>\>\biggl ( s_5+s_3+s_1-c (s_4+s_2+1) -c^2 (s_3+2 s_1)+ \nonumber \\ &&+c^3 (s_2+2) +c^4 s_1-c^5\biggr ) 
 \biggl (s_4+s_2+1-c(s_5+s_3+s_1)- \\
&&-c^2 (s_4+2 s_2+3) +c^3 (s_3+2 s_1)+c^4 (s_2+3) -c^5 s_1-c^6 \biggr )^{-1}, \nonumber
\end{eqnarray}
where we have used short notations for the symmetric functions 
\begin{eqnarray}
\label{symu}
&&s_1=\cos 2u_1+\cos 2u_2 +\cdots +\cos 2u_5 \nonumber \\
&&s_2=\cos 2u_1 \cos 2u_2 +\cdots +\cos 2u_4 \cos 2u_5 \nonumber \\
&& \dots \\
&&s_5=\cos 2u_1 \cos 2u_2 \cdots \cos 2u_5 \nonumber
\end{eqnarray}
and for the trigonometric functions: $c=\cos 2u$ and $s=\sin 2u$.
It is clear that 
 \begin{eqnarray}
\label{apident}
&& \prod_{j=1}^{5} (\cos 2u_j - \cos 2u) = s_5-c\>s_4+c^2\>s_3-c^3\>s_2+c^4\>s_1-c^5=0,  
\end{eqnarray}
for $u=u_i$.
Using the last identity we can exclude $s_5$ from (\ref{left2}).
After some simplifications 
 we obtain the more simple fraction:
  \begin{eqnarray}
\label{left3}
&&\frac{L(u)-1}{L(u)+1}=s \frac{s_3+s_1-c (s_2+1)}
{s_4+s_2+1-c(s_3+s_1)-c^2+c^4},\>\> u=u_1,\ldots,u_5. 
\end{eqnarray}
 
 On the other hand the right side of the  third group of equations (\ref{apbae3}) can be written as 
$R(u_i),\>\>i=1,\ldots,5$, where 
 \begin{eqnarray}
\label{right1}
&&R(u)=\prod_{j=1,}^{3} f(u,v_j). 
\end{eqnarray}
Again, rather straightforwardly  we obtain:
  \begin{eqnarray}
\label{right2}
&&\frac{R(u)-1}{R(u)+1}=-s \frac{2\sqrt{2} \sigma_2+4c\>\sigma_1+\sqrt{2}(1+2c^2)}
{4\sigma_3+2\sigma_1+2\sqrt{2} c\>\sigma_2+\sqrt{2}(3c-2c^3)},
\end{eqnarray}
where we have used the  short notations for the symmetric functions 
\begin{eqnarray}
\label{symv}
&&\sigma_1=\cos 2v_1+\cos 2v_2 +\cos 2v_3 ,\nonumber \\
&&\sigma_2=\cos 2v_1 \cos 2v_2 +\cos 2v_1 \cos 2v_3+\cos 2v_2 \cos 2v_3, \\
&&\sigma_3=\cos 2v_1 \cos 2v_2 \cos 2v_3. \nonumber
\end{eqnarray}
 By equating the right sides of equations 
(\ref{left3}) and  (\ref{right2}) 
we obtain a polynomial equation of degree 6  with respect to the 
variable $c$.
Using (\ref{apident}) we can exclude the terms with $c^6$ 
and $c^5$  and obtain 
a polynomial equation of the 4th  degree with 
respect to the variable $c$.
It is amusing to verify that all the five  coefficients of this 
polynomial become zero if we choose  
\begin{eqnarray}
\label{resultingv}
&&\sigma_1=-\frac{2s_1+s_3+s_1\>s_2}{\sqrt{2}(1+s_2)},\nonumber \\
&&\sigma_2= -\frac{1}{2}+\frac{s_1(s_1+s_3)}{1+s_2},\\
&&\sigma_3=\frac{s_3+2s_5-s_1\>s_2-2s_1\>s_4}{2\sqrt{2}(1+s_2)}. \nonumber
\end{eqnarray}
This result means that for 
 arbitrary values of $u_1,\ldots,u_5$ we calculate 
by using (\ref{symu}) $s_1,\ldots,s_5$,
and the above relations (\ref{resultingv}) together with 
the definition 
(\ref{symv}) enable us to find the variables 
$v_1,v_2,v_3$ as the roots of the equation
\begin{eqnarray}
\label{cube}
&&\cos^32v-\sigma_1 \cos^22v +\sigma_2 \cos 2v -\sigma_3=0 \quad 
(v=v_1,v_2,v_3).
\end{eqnarray}

Let us consider now the first  equation:
 \begin{eqnarray}
\label{apfirst}
&&\prod_{j=1,}^{3} f(w,v_j)-1 \equiv R(w)-1 = 0, 
\end{eqnarray}
where $R$ is given by 
(\ref{right1}). 
Due to  (\ref{right2}) the equation (\ref{apfirst}) 
becomes:
  \begin{equation}
\label{findw1}
2\sqrt{2} \sigma_2+4\cos 2w\>\sigma_1+\sqrt{2}(1+2\cos^2 2w)=0.
\end{equation}
Inserting (\ref{resultingv}) into this last equation we obtain 
\begin{equation}
\label{findw2}
(s_1-\cos 2w) \biggl(\frac{s_1+s_3}{1+s_2}-\cos 2w \biggr)=0.
\end{equation}
We have two possible solutions  for $\cos 2w$.
Choosing the solution 
\begin{equation}
\label{apresw}
\cos 2w= \frac{s_1+s_3}{1+s_2},
\end{equation}
 we can check by using standard algebra that for this 
choice of $w$ and $v_1,v_2,v_3$ obtained  
from  (\ref{cube}) all equations of  
second group (\ref{apbae2}) are satisfied automatically! 

As a result, we explained  one more free-fermion branch found in 
our numerical observations.

\begin{center}{\bf Appendix B. Eigenvectors for the eigenvector in the 
sector ($1,\ldots,1,L-N+1$)}
\end{center}

In \S 4 we found a set of solutions in the sector ($1,\ldots,1,L-N+1$)
of the $SU(N)_q$ model with an arbitrary value of the anisotropy $q$, 
and free boundary conditions.
The corresponding eigenenergies are given by (\ref{ba1f}), and can be 
written as:
\begin{equation} \label{apb1}
E = \sum_{j=1}^{N-1} \epsilon_j = \sum_{j=1}^{N-1} (q +\frac{1}{q} -x_j 
-\frac{1}{x_j}),
\end{equation}
where
\begin{equation} \label{apb2}
x_j = \exp(i\frac{\pi k_j}{L}), \quad 1\leq k_j \leq L-1.
\end{equation}

We intend to prove that the corresponding eigenvectors have the following 
form
\begin{eqnarray} \label{apb3}
&&|\psi_{\{x_1,\ldots,x_{N-1}\}}> = \sum_{m_1,\ldots,m_{N-1}= 1}^L
q^{-f(m_1,\ldots,m_{N-1})} \nonumber \\
&& \times \mbox{det}
\left|\begin{array}{cccc} 
\Psi_1(m_1) & \Psi_1(m_2) & \cdots & \Psi_1(m_{N-1}) \\
\Psi_2(m_1) & \Psi_2(m_2) & \cdots & \Psi_2(m_{N-1}) \\
\cdots      & \cdots      & \cdots & \cdots          \\
\Psi_{N-1}(m_1) & \Psi_{N-1}(m_2) & \cdots & \Psi_{N-1}(m_{N-1}) 
\end{array} \right|
|m_1,\ldots,m_{N-1}>, \nonumber
\end{eqnarray}
where $|m_1,\ldots,m_{N-1}>$ are the vector basis representing the 
configuration where the   $i$th particle 
is located at site $m_i$ ($i=1,\ldots,N-1$, $1 \leq m_i \leq L$). The Slater determinants 
entering in (\ref{apb3}) depend upon the set of one-particle amplitudes 
$\Psi_j(m)$, that apart from a normalization factor are given by 
\begin{equation} \label{apb4}
\Psi_j(m) = \left( 1-\frac{q}{x_j}\right)x_j^m - \left(1-qx_j\right)/x_j^m.
\end{equation}
The factor $f(m_1,\ldots,m_{N-1})$ is an integer number that depends on the 
relative order of the sequence $\{m_1,m_2,\ldots,m_{N-1}\}$. We are going to find 
this dependence in the procedure of proving (\ref{apb3}).

The application of the Hamiltonian (\ref{H}) to the above vector leads 
to a set of equations for the amplitudes 
$\Psi(m_1,\ldots,m_{N-1})$. 

Consider initially the amplitudes where 
$|m_i - m_j| \geq 2$ for all pairs ($i,j$). These 
amplitudes 
give us a  set of equations ("regular" ones) 
for one-particle amplitudes that are  satisfied for 
every product in the determinant separately.
 Consider now the case where $m_i =m$ and $m_j = m+1$ for precisely a 
single pair ($i\leq j$). The equation in this case are given by
\begin{eqnarray} \label{apb5}
&&E\Psi(\ldots,m,\ldots,m+1,\ldots) = -\sum_{k=1,k\neq(i,j)}^{N-1} 
\left[\Psi(\ldots,m_k-1,\ldots) \right.\nonumber \\
&&\left.+ \Psi(\ldots,m_k+1,\ldots)\right] 
-\Psi(\ldots,m_1,\ldots,m+1,\ldots) \nonumber \\
&&-\Psi(\ldots,m\ldots,m+2) 
-\Psi(\ldots,m+1,\ldots,m,\ldots) \nonumber \\
&&+\left[(N-1)q +(N-2)/q\right] \Psi(\ldots,m,\ldots,m+1,\ldots).
\end{eqnarray}
Comparing this equation with the "regular" ones, derived previously we obtain
\begin{eqnarray} \label{apb6}
&&\Psi(\ldots,m,\ldots,m,\ldots) + \Psi(\ldots,m+1,\ldots,m+1,\ldots)
\nonumber \\ 
&& -\Psi(\ldots,m+1,\ldots,m,\ldots) - \frac{1}{q}\Psi(\ldots,m,\ldots,m+1,\ldots) 
= 0
\end{eqnarray}
Inserting (\ref{apb3}) into (\ref{apb6}) we verify that the first two 
terms vanishes while the third and fourth ones correspond to the same 
determinant, apart from a minus sign. As a result we obtain an equation for 
the factor $f$ in (\ref{apb3}):
\begin{equation} \label{apb7}
f(\ldots,m,\ldots,m+1,\ldots) = -1 + f(\ldots,m+1,\ldots,m,\ldots).
\end{equation}
We are now ready to fix the function $f$.
Let $f(m_1,m_2,\ldots,m_{N-1}) =0$, for all $\{m_j\}$ satisfying to the  
order:
\begin{equation} \label{apb8}
m_1 <m_2 < \cdots <m_{N-1}.
\end{equation}
Then (\ref{apb7}) fixes $f$ for all remaining configurations.

	Consider, for example, the $SU(4)_q$ case. We have 6 possibilities 
\begin{eqnarray} \label{apb9}
&&f(m_1 <m_2<m_3) = 0 \nonumber \\
&& f(m_2<m_1<m_3) = f(m_1<m_3<m_2) = 1 \nonumber \\
&&f(m_2<m_3<m_1) =f(m_3<m_1<m_2) = 2 \nonumber \\
&&f(m_3<m_2<m_1) = 3. \nonumber 
\end{eqnarray}
It is clear that $f$ is equal to the minimal number of pair transpositions 
necessary to put the configuration in the order $m_1<m_2<m_3$.

One can easily check that the remaining  equations coming from the other 
 amplitudes reduce to the already obtained relation (\ref{apb7}), concluding the 
proof that (\ref{apb3}) are the eigenfunctions corresponding to the NBAE  solutions 
derived in \S 4, with eigenvalues given by (\ref{apb1}).

\end{document}